# Low Temperature Conduction Behavior in Highly Crystallized Undoped Microcrystalline Silicon Thin Films

Sanjay K. Ram,*,1 Satyendra Kumar*,2 and P. Roca i Cabarrocas!

*Department of Physics, Indian Institute of Technology Kanpur, Kanpur-208016, India
!LPICM, UMR 7647 - CNRS - Ecole Polytechnique, 91128 Palaiseau Cedex, France

The temperature dependence of dark conductivity $\sigma_d(T)$ at low temperatures (300-15 K) was studied on a wide microstructural range of well-characterized highly crystallized single phase undoped microcrystalline silicon (µc-Si:H) samples. Our study reveals two different temperature dependences in films having different microstructures. A $T^{-1/2}$ dependence of $\sigma_d(T)$ supporting tunneling of carriers between neighboring conducting crystals, similar to percolation-hopping model proposed for metal-insulator composite systems, is seen in µc-Si:H films that are fully crystallized with tightly packed large columnar grains and negligible density deficit. A $T^{-1/4}$ dependence of $\sigma_d(T)$ supporting variable range hopping model with an exponential tail state distribution in the gap is seen in µc-Si:H films having mostly small crystalline grains, low degree of conglomeration and relatively higher density deficit. The correlation between the microstructural attributes and conductivity behavior is discussed by analyzing the physical plausibility of the hopping parameters and material properties derived by applying different transport models.

## I. INTRODUCTION

Hydrogenated microcrystalline silicon (µc-Si:H) has many attractive properties like good mobilities,[1] stability against light degradation[2,3] and the possibility of large area deposition which are advantageous for applications in photovoltaics[4] and thin film transistors.[5,6] Nevertheless, the complicated heterogeneous microstructure of µc-Si:H system and the lack of knowledge about the distribution of density of states (DOS) in it have not permitted an adequate understanding of this material even after years of research. In a disordered system, carriers can be thermally activated across mobility edges into the extended states giving rise to an Arrhenius type temperature dependence in the conductivity. In general, electrical transport in µc-Si:H films shows an activated behavior,[7] although different transport mechanisms may be observed depending on the film microstructure, doping and the range of measurement temperatures. Electrical transport at low temperature measurements has been explained in terms of tunneling[8] and hopping.[9,10,11,12,13] In addition to these models, considering the difference in conductivities of crystalline and amorphous phases in µc-Si:H system, a tunneling mechanism similar to that observed for metal-insulator composite systems has also been suggested.[14,15,16,17]

Because µc-Si:H is not a microstructurally defined unique material, a certain model that satisfactorily explains electronic transport for a particular µc-Si:H material may not be applicable to the whole system. For the elucidation of a transport mechanism applicable to a wider range of µc-Si:H materials having different and distinct microstructural and morphological attributes, a correspondingly wide range of µc-Si:H films have to be studied. While applying any model to a limited amount of data in the face of the numerous models suggested in literature, the plausibility of the mechanism has to be ascertained in context of the various physical parameters. Keeping these issues in mind, in this article we have presented the results of dark conductivity measurements conducted below room temperature (300 to 15 K) on a wide range of well-characterized highly crystallized single phase undoped µc-Si:H samples,[18] addressing the current transport mechanisms in the context of the complex microstructure.

The outline of the article is as follows. In the next section, the various low temperature electrical transport mechanisms are discussed, in the context of disordered systems. In Sec. III, the experimental details of our work are described, along with a brief outline of the findings of the structural studies of our material. In Sec. IV, the results of low temperature conductivity measurements are presented, followed by a discussion of the results in Sec. V. The findings of this study are summarized in Sec. VI.

## II. LOW TEMPERATURE CONDUCTION MECHANISMS

The temperature dependence of dark carrier conduction $\sigma_d$ is given by:

$$\ln(\sigma_d) \propto (T_0/T)^m \qquad (1)$$

where $T_0$ is a constant for the material.[19] The value of $m$ depends on the grain sizes and the temperature region and gives information on the type of carrier conduction mechanism. In a bulk material, $m= ¼$ is attributed to Mott's VRH conduction (M-VRH) where electron-electron ($e$–$e$) interactions are neglected.[19] According to Mott's equation for the three-dimensional case, if the DOS is constant in a $kT$ energy range around Fermi level ($E_f$), the $\sigma_d(T)$ is expressed as:

$$\sigma_d = \sigma_0^* \exp(-T_M/T)^{¼} \qquad (2)$$

where the terms $\sigma_0^*$ and $T_M$ are constants. $T_M$ is related to the DOS at the Fermi energy ($N_f$) by expression: $T_M = C_M \alpha^3/(kN_f)$, where $k$ is Boltzmann's constant, $\alpha$ is the decay constant of localized wave function and $C_M$ is a constant. According to M-VRH, the hopping parameters, optimum hopping distance ($R_{opt}$) and optimum hopping energy ($W_{opt}$) are expressed as:

$$W_{opt} = \frac{3}{4\pi R^3 N_f} \qquad (3)$$

and

$$R_{opt} = \left[\frac{9}{8\pi \alpha N_f kT}\right]^{¼} \qquad (4)$$

---

[1] Corresponding author. E-mail address: skram@iitk.ac.in, sanjayk.ram@gmail.com
[2] satyen@iitk.ac.in

The average hopping distance is ¾ of $R_{opt}$. The criteria for M-VRH to be valid include a basic initial assumption of a constant DOS around the $E_f$, $\alpha R_{opt}(T) \geq 1$, and that the jump of the carriers from one state to the other is only of the order of some $kT$. Variation in the value of $\alpha^{-1}$ has been reported between 10-20 Å for amorphous material,[20,21] whereas large value of $\alpha^{-1} \approx 257$ Å has been reported for polycrystalline material.[22] Variation in the value of $C_M$ has been a more contentious issue.[23] According to Mott,[19] the value of $C_M$ is $\approx 16$ for a constant DOS around the $E_f$. However, Godet[24,25] reported high values of $C_M$ (=310) for materials having an exponentially distributed DOS (G-VRH model) and set two conditions for this to be applicable. First, a linear relationship should be present between $\ln \sigma_0^*$ and the slope $T_M^{¼}$, and secondly, the localization parameter (LP =$N_f.\alpha^{-1}$) should be in the range of $10^{-5}$ -1. Due to the variation in $C_M$, different approximations used to express VRH parameters in terms of $T_M$ often lead to contradictory results, more so in $W_{opt}$. This has sometimes resulted in the M-VRH model being challenged, and alternative models being suggested as options that are more plausible.[13] $R_{opt}$ shows less variation in results of such approximations because the factor $\alpha^{-1}$ ($\approx 10^{-7}$ cm) minifies any variation induced by a change in the value of $C_M$. A similar situation is evident in the diffusion model given by Dussan et al.,[13] where the original equations of M-VRH expressed in terms of $N_f$ have been used for hopping parameters, but they have been inappropriately equated to the equations derived in other works[12,22] that had chosen a specific value for $C_M$ to express M-VRH equations in terms of slope $T_M$. In fact the equations derived for the diffusion model are the very same Mott's equations, and in one of the coauthor's previous work[11,12] on similar p-type Boron doped $\mu$c-Si:H:(B) and intrinsic $\mu$c-Si:H material, G-VRH model was found to be valid and hopping parameters were deduced for M-VRH. The equations described as classical VRH equations in Ref. [13] are in reality derivations of the basic M-VRH equations that have been approximated using specific constant values for $C_M$ [$16\times(3/2\pi)$ =7.64]. The constant terms associated with $N_f$ cannot be ignored when $N_f$ is incorporated in the equations for hopping parameters (Eqs. 3 & 4).

Therefore, we have not considered the diffusion model for the choice of hopping parameters. Instead, in our opinion, it is more appropriate to use the equations given by Rosenbaum et al., according to which, $R_{opt}$ and $W_{opt}$ in terms of $T_M$ and $C_M$ can be expressed as:[23]

$$R_{opt} = \frac{1}{\alpha}\left[\frac{3^{½}}{2^{¾}\pi^{¼}C_M^{¼}}\right]\left(\frac{T_M}{T}\right)^{¼} \qquad (5)$$

and

$$W_{opt} = k\left[\frac{2^{¼}}{3^{½}\pi^{¼}C_M^{¼}}\right](T_M T^3)^{¼} \qquad (6)$$

A $m=½$ temperature dependence of carrier transport may arise either due to Efros-Shklovskii variable range hopping (ES-VRH)[26] or due to percolation-hopping transport model as applicable to a composite system of granular metal dispersed in an insulator matrix proposed by Šimánek.[27]

If we include the long-range $e$–$e$ interactions (i.e., Coulomb interaction between the localized state electrons) which cause the DOS to tend to zero in the vicinity of $E_f$ resulting in a soft energy gap (Coulomb gap) corresponding to the parabolic DOS, we find that carrier transport due to hopping in the Coulomb gap in DOS near the Fermi energy exhibits a $m=½$ temperature dependence as: $\ln(\sigma_d) \propto (T_0/T)^{1/2}$, which is known as ES-VRH.[26] In ES hopping, constant $T_0$ is related to Coulomb gap energy ($\Delta$) and DOS ($g_o$) as:[28]

$$\Delta = (k/2)(T_0 T_r)^{1/2} \qquad (7)$$

and

$$g_0 = 3^8 \pi^2 k^2 \varepsilon_r^3 \varepsilon_o^3 /(2^9 q^6)(T_0 T_r) \qquad (8)$$

Here, $\varepsilon_r$ is the relative dielectric constant, $q$ the electronic charge and $T_r$ the transition temperature between Mott and ES hopping conduction.[26,28,29,30]

According to percolation-hopping model proposed by Šimánek, in a metal-insulator composite system, the carriers traversing from one grain to another grain face a barrier height $E_b$, given by:[27,28,31,32]

$$E_b = 2se^2/\{\varepsilon d_g(s+d_g/2)\} \qquad (9)$$

where $d_g$ is the average grain size, $s$ is the average width of the high resistive region between the neighboring grains, $\varepsilon$ is the dielectric constant of the insulator and $e$ is the electronic charge. At low electric field when the applied voltage drop ($e\Delta V$) across the two nearest neighboring conducting regions is less than $E_b$, as well as less than $kT$, thermal assisted tunneling occurs, which leads to:

$\ln(\sigma_d) \propto (T_0/T)^{1/2}$. The value of $T_0$ is given by:[31]

$$T_0 = 16 \chi e^2 \left(\frac{s}{d_g}\right)^2 \bigg/ k\varepsilon \left(\frac{1}{2} + \frac{s}{d_g}\right) \qquad (10)$$

where $\chi$ is the rate of decay of the wave function in the insulator.

## III. EXPERIMENTAL DETAILS

Undoped $\mu$c-Si:H films were deposited at a low substrate temperature ($T_s \leq 200°C$) in a parallel-plate glow discharge plasma enhanced chemical vapor deposition system operating at a standard rf frequency of 13.56 MHz, using high purity $SiF_4$, Ar and $H_2$ as feed gases. The microstructural studies were carried out employing bifacial Raman scattering (RS), spectroscopic ellipsometry (SE), X-ray diffraction (XRD), and atomic force microscopy (AFM). The findings of the structural studies are briefly as follows. High crystallinity of all the samples was confirmed by RS and SE measurements. SE data demonstrated a crystalline volume fraction >90% in the bulk of the material from the initial stages of growth, with the rest being density deficit having no amorphous phase. However, a very thin ($\approx$20 nm) film-substrate interface incubation layer is present, which is less crystallized and contains some amorphous content. The fractional composition of the films educed from SE data revealed crystallite grains of two distinct sizes,[33,34] which was corroborated by the deconvolution of RS profiles using a bimodal size distribution of large crystallite grains (LG $\approx$70–80nm) and small crystallite grains (SG $\approx$6–7nm).[35,36] The XRD results have correspondingly demonstrated the LG and SG to be having different orientations. There is a significant variation in the percentage fraction of the constituent LG and SG with film growth. Although the concept of a bimodal distribution in the sizes of crystallites is fundamental to the application of percentage fraction of large crystallites as a microstructural parameter, but the details of this modeling method are beyond the scope of the present paper, and are being reported elsewhere. The evolution of conglomeration of crystallites with film growth, under a variety of deposition conditions was demonstrated by AFM results. Preferential orientation in (400) and (220) directions as achieved by optimizing the deposition conditions leading to smooth top surfaces (surface roughness < 3nm), indicating device quality material.[18,33,34] The dark conductivity $\sigma_d(T)$ measurements were carried out on a large number of well-annealed samples having different thicknesses, microstructures and morphological properties, using coplanar geometry in different experimental set-ups (above room temperature, 300–450K; and low temperature, 300–15 K).

Based on the structural investigations of the $\mu$c-Si:H films at various stages of growth and under different growth conditions,[18,34] we segregated out the unique features of microstructure and growth type present in the varieties of $\mu$c-Si:H films, with respect to the correlative coplanar electrical transport properties (above room temperature). The results of $\sigma_d(T)$ measurements above room temperature are not presented here per se,[37] but are mentioned in context of the classification of the films. Obviously, deposition parameters merely have an indirect contributory role in the electrical properties through their fundamental effect on the microstructure of material. Our structural and electrical studies[18,34] have indicated that the fractional composition of large constituent crystallite grains ($F_{cl}$, determined by SE) is a simple yet physically rational microstructural parameter that indicates the microstructural and morphological condition of the highly crystallized single phase undoped $\mu$c-Si:H film and correlates well with the electrical transport behavior at high temperatures.[18,37] All the $\mu$c-Si:H samples were classified into three types: *A, B* and *C*, where we see the influence of the nature of inhomogeneities in $\mu$c-Si:H not only on the $\sigma_d(T)$ but also on the material parameters like dark conductivity prefactor ($\sigma_0$) and $E_a$.

To summarize this classification, the *type-A* films have small grains, high density of inter-grain boundary regions containing disordered phase, and low amount of conglomeration. In this type, $F_{cl}$ <30%, $\sigma_0$ and $E_a$ are constant [$\approx 10^3$ $(\Omega cm)^{-1}$ and $\approx$0.55 eV respectively]. The *type-B* films contain a fixed ratio of mixed grains in the bulk. With film growth, conglomeration of grains results in a marked morphological variation, and a moderate amount of disordered phase in the conglomerate boundaries limits the electrical transport. In this type, $F_{cl}$ varies from 30% to 45%, there is a sharp drop in $\sigma_0$ [from $\approx 10^3$ to 0.1 $(\Omega cm)^{-1}$] and $E_a$ (from $\approx$0.55 to 0.2 eV). The *type-C* $\mu$c-Si:H material is fully crystallized and crystallite conglomerates are densely packed with significant fraction of large crystallites (>50%) and preferential orientation is seen. In this material, $\sigma_0$ shows a rising trend [from 0.05 to 1 $(\Omega cm)^{-1}$] and the fall in $E_a$ is slowed down (from 0.2 to 0.10 eV).

## IV. RESULTS

Since it is not possible to report all the results of the large number of samples studied, for the sake of conciseness, we have presented the results of a few representative samples from each type in this article. These samples are listed below with the respective film thickness and $R$ (=$SiF_4/H_2$) values. *Type-A* samples: #B22 (170nm, $R$=1/10) and #D26 (410nm, $R$=1/5); *type-B*: #B11 (390nm, $R$=1/10), #B23 (590nm, $R$=1/10) and #B04 (950nm, $R$=1/10); *type-C*: #F06 (920nm, $R$=1/1), #E31 (1200nm, $R$=1/1) and #E25 (1025nm, $R$=1/5). Coming to the results of the low temperature measurements, Fig. 1 shows $\sigma_d(T)$ plotted with reciprocal of $T$. A majority of samples showed single activation energy from $\approx$450K to $\approx$200K, indicating excellent material properties from device application point of view. However, a careful analysis of low temperature data reveals a continuously vary-ing slope in $\sigma_d(T)$, which becomes almost independent of temperature for all samples at temperatures lower than certain values, depending on the film microstructure. In Fig. 2, we have plotted the low temperature $\sigma_d(T)$ data versus $T^{-1/4}$ according to M-VRH. At low temperatures, the linear fitting seems to be good with reasonable standard deviation values. Using M-VRH model the slopes $T_M$ were determined from the graph and are listed in Table I. The values of $N_f$, $R_{opt}$ and $W_{opt}$ calculated from the slopes $T_M$ for both the conditions, i.e., hopping near $E_f$ (M-VRH, constant DOS around $E_f$, $C_M=18.1$) and hopping in bandtail states (G-VRH, exponential DOS, $C_M=310$) for $\alpha^{-1}=14$Å are presented in Table I. Although the prerequisite condition for M-VRH ($\alpha R_{opt}(T) \geq 1$) is fulfilled here, but if we consider the conditions set out by Godet to discriminate between hopping near $E_f$ and hopping in bandtails, we see that a linear relationship is pre-sent between $\ln \sigma_0^*$ and the slope $T_M^{1/4}$ (shown in the inset of Fig. 2) and the values of LP are lying well within the range of $10^{-5} - 1$, which suggests hopping in our $\mu$c-Si:H material is taking place in exponentially distributed bandtail states. The hopping parameters will be discussed a little further on.

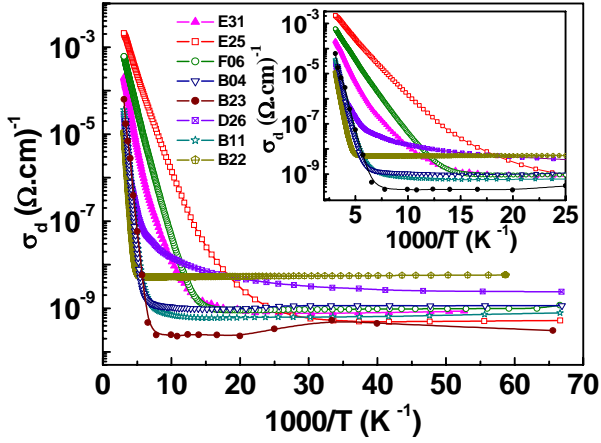

FIG. 1. Temperature dependence of $\sigma_d(T)$ at low temperatures. The sample details are: *Type-A* samples: #B22 (170nm, $R=1/10$) and #D26 (410nm, $R=1/5$); *type-B*: #B11 (390nm, $R=1/10$), #B23 (590nm, $R=1/10$) and #B04 (950nm, $R=1/10$); *type-C*: #F06 (920nm, $R=1/1$), #E31 (1200nm, $R=1/1$) and #E25 (1025nm, $R=1/5$). $T_s$ of all the samples is 200°C.

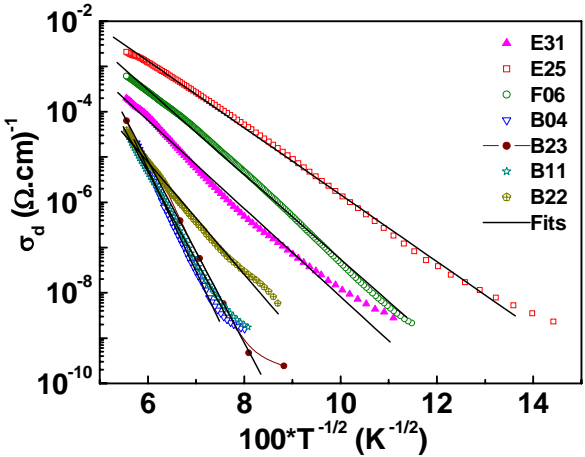

FIG. 2. Low temperature $\sigma_d(T)$ data as a function of $T^{-1/2}$.

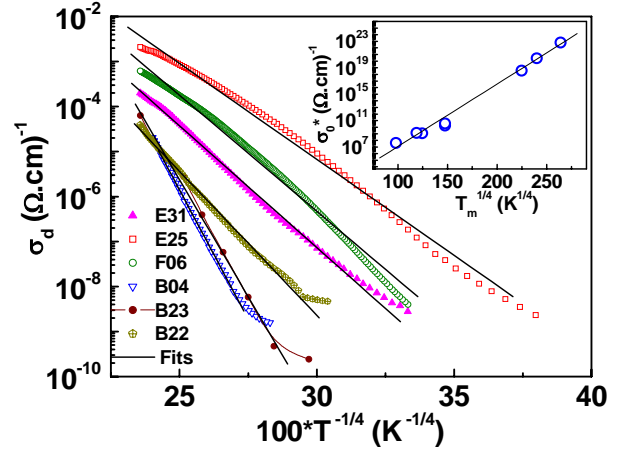

FIG. 3. Low temperature, $\sigma_d(T)$ data as a function of $T^{-1/4}$. The inset shows relationship between $\sigma_0^*$ and slope $T_M^{1/4}$.

Although our low temperature $\sigma_d(T)$ data seems to be fitted well with the VRH model, it is still desirable to see if any other type of fitting is possible. A fitting of the same low temperature data to the conductivity expression, $\sigma_d \approx exp\ (T_0/T)^{1/2}$, is demonstrated in Fig. 3. Here $T_0$ is a constant for the material. Value of $T_0$ is found to decrease from $2.7\times10^5$ to $3\times10^4$ K with increasing film thickness or with the improvement in film microstructure. The unrealistically large Coulomb gap obtained from $T_0$ does not permit ES-VRH mechanism to be applied in this temperature range and hence can be ruled out.

In the heterogeneous structure of our $\mu$c-Si:H samples, the granular bulk (crystalline columnar island) conductivity is larger than the conductivity in the boundary regions containing disordered phase. The overall conductivity is governed by the resistive inter-granular region. Therefore, we have considered the applicability of the percolation-hopping model proposed by Šimánek as well. The values of $T_0$ and $s/d_g$ obtained (using the value of $\chi \approx 1$ Å$^{-1}$, and $\varepsilon \approx 8$ for bulk-disordered region) for various samples are listed in Table I. It is interesting to note that the temperature range over which $T^{-1/2}$ behavior can be fitted increases with film thickness and/or improving film microstructure. The values of $s/d_g$ also show the right trend as expected from the film microstructure, which is further discussed in the next section. Similar temperature dependent $\sigma_d(T)$ behavior is also reported for SiF$_4$ based plasma deposited fully crystallized $\mu$c-Si:H[16] and recently in Si nanocrystals[17].

## V. DISCUSSION

If we consider the VRH mechanisms, the G-VRH model seems to be applicable to our observations, as discussed above. The reason behind this maybe the fact that the improvement in film microstructure leads to delocalization of the tail states causing the $E_f$ to move towards the band edges. In contrast to the constant DOS at $E_f$ in amorphous silicon, the $E_f$ in $\mu$c-Si:H is actually somewhere in the vicinity of the exponential tail states. Therefore, in $\mu$c-Si:H system, both doped and intrinsic, the G-VRH mechanism is more applicable than the M-VRH. Recently, similar observation was also reported by Myong et al.[38,39] for hydrogenated boron (B)-doped nc-Si-SiC:H (p-nc-Si-SiC:H alloy) material and for various kinds of inorganic semiconducting polycrystalline films.[40] It is to be noted that the values we obtained for $N_f$ for M-VRH in Table I are one order less than the values obtained for G-VRH for all the samples, an observation similar to what has been reported for doped and intrinsic $\mu$c-Si:H.[11] This is because the term $N_f$ in G-VRH is actually the DOS with an exponential distribution near $E_f$. The $R_{opt}$ and $W_{opt}$ values are also lower in G-VRH compared to M-VRH. $R_{opt}$ is actually the nearest neighbor hopping distance of a carrier as it traverses from one state to another. In case of a constant DOS, any two 'nearest neighbor' states may have only a small amount of energetic separation, but a larger actual interatomic distance between them. In case of hopping in a bandtail where the states are exponentially distributed, the hopping distance is reduced. Similar reasoning applies to $W_{opt}$, because in an exponential distribution of states, the nearest neighboring states have reduced energetic separation.

Let us now consider the hopping parameters obtained from VRH mechanism ($T^{-1/4}$ dependence) and material properties deduced from $T^{-1/2}$ dependence. It is desirable to see if these hopping parameters and material properties are corroborative with the film microstructure. The variation of $N_f$ as a continuum with film thickness is shown in Fig. 4(a), and the same data points are shown as discrete values corresponding to the respective type of the material in Fig. 4(b). We have shown the variation of $N_f$ with percentage volume fraction of large crystallites ($F_{cl}$) in Fig. 4(c), where, on proceeding from type-A to type-B material, we note a gradual fall in $N_f$ up to one order. The trend reverses and $N_f$ starts rising at about $F_{cl} \approx 40$ reaching a plateau in type-C above $F_{cl} \approx 50$. If we see the variation of $N_f$ with thickness in Fig. 4(a), an erratic relation is visible. This zigzag appearance of the graph actually occurs due to the erroneous attempt to correlate a material property ($N_f$) with increasing thickness, and gets eliminated in Fig. 4(b), where the $N_f$ values are more appropriately considered in context of the film type (microstructure) and the data points can be now seen as clusters in three regions. The observed variation of $N_f$ with film microstructure is explained by the changes occurring in the DOS with film growth.[41,42,43] At the beginning of the growth, the DOS is high, and in spite of being fully crystalline, the film is mainly composed of small grains, with a relatively high number of grain boundaries resulting in high defect densities. With film growth, the $F_{cl}$ rises, there is alteration in film morphology with the onset of conglomeration of grains, and the number of grain boundaries are reduced, resulting in a steep fall in the $N_f$ by an order of magnitude. Further, as we approach *type-C* material, $N_f$ rises, as the $E_f$ gets closer to conduction band (CB) edge, which is steeper compared to the CB edge in *type-A* and *type-B* materials. Another possibility is the unintentional impurities distributed near CB edge, which can also enhance the overall DOS in the vicinity of $E_f$. Here, we observe that it is the film microstructure parameterized by $F_{cl}$ (rather than film thickness) that plays a crucial role in determining the electrical transport behavior of the material. This is the basis of our classification discussed in Sec. III, where the $F_{cl}$ has been correlated to the electrical transport properties. While it is tempting to correlate the classification to thickness, as each type of $\mu$c-Si:H material apparently seems to extend over a certain range of thickness as well, but this can be misleading, as we found in all our electrical transport studies, including this study.

Table I. The physical parameters, viz., DOS values and hopping parameter values ($T$=200K) obtained by application of various fitting procedures and models to low temperature conductivity data.

| | | $T^{-1/2}$ dependence | | $T^{-1/4}$ dependence | | | | | | | |
|---|---|---|---|---|---|---|---|---|---|---|---|
| | | | | | Godet's Model (G-VRH) | | | | Mott's Model (M-VRH) | | |
| Sample# | $F_{cl}$% | $T_0$(K) ($10^4$) (fitting T range) | $\sim s/d_g$ | $T_M$(K) ($10^8$) | $N_f$ (eV$^{-1}$cm$^{-3}$) ($10^{17}$) | $R_{opt}$ (nm) | $W_{opt}$ (mev) | LP(eV$^{-1}$) ($10^{-3}$) | $N_f$ (eV$^{-1}$cm$^{-3}$) ($10^{16}$) | $R_{opt}$ (nm) | $W_{opt}$ (meV) |
| E31 | 90 | 5.02 (90 - 325K) | 0.072 | 2.41 | 54.5 | 8.55 | 70.17 | 15 | 31.8 | 17.4 | 142.8 |
| E25 | 55.8 | 2.89 (60 - 325K) | 0.05 | 0.93 | 141 | 6.74 | 55.33 | 39 | 82.3 | 13.7 | 112.6 |
| F06 | 47.5 | 4.65 (85 - 325K) | 0.07 | 1.99 | 66 | 8.15 | 66.91 | 18 | 38.5 | 16.6 | 136.1 |
| B04 | 40.5 | 27.1 (186 - 300K) | 0.18 | 48.6 | 2.7 | 18.1 | 148.7 | 0.74 | 1.6 | 36.9 | 302.5 |
| B23 | 35.6 | 21.4 (150 - 325K) | 0.16 | 33.2 | 3.9 | 16.5 | 135.23 | 1.1 | 2.3 | 33.5 | 275.1 |
| B11 | 31.5 | 20.2 (150 - 325K) | 0.15 | 25.5 | 5.1 | 15.4 | 126.61 | 1.4 | 3.0 | 31.4 | 257.6 |
| D26 | 4.75 | 7.82 (174 - 325K) | 0.09 | 4.74 | 27.7 | 10.13 | 83.11 | 7.6 | 16.15 | 20.6 | 169.1 |
| B22 | 8.85 | 8.14 (180 - 325K) | 0.09 | 4.74 | 1.62 | 10.1 | 80 | 7.6 | 16.17 | 20.6 | 169 |

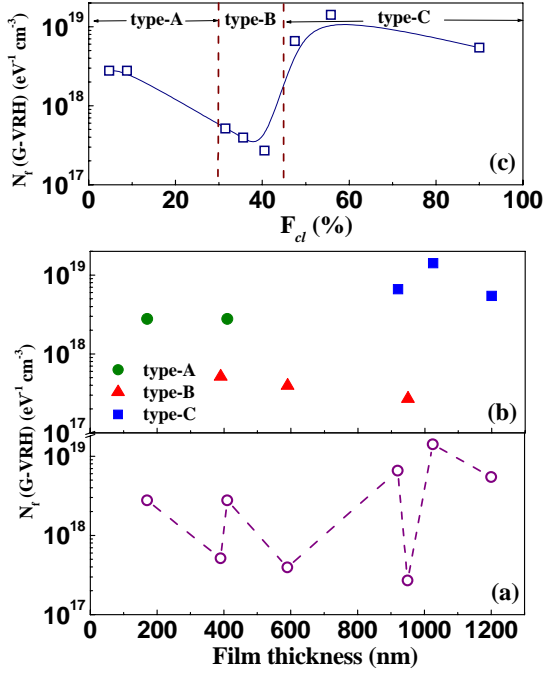

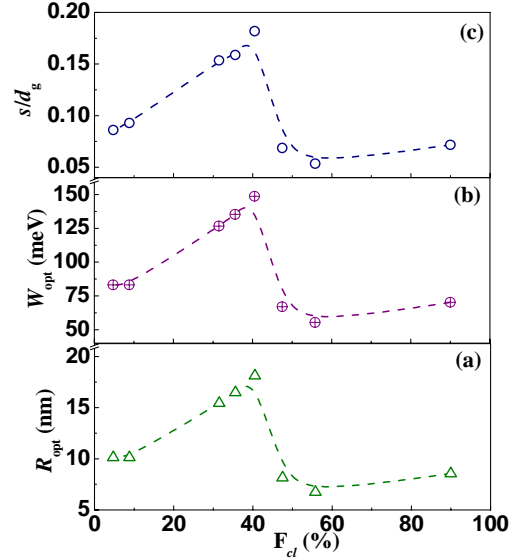

FIG. 4. (a) Variation of $N_f$ as a continuum with film thickness; (b) the same variation of $N_f$ shown in context of individual types (*A*, *B* and *C*) of material; (c) variation of $N_f$ with $F_{cl}$.

FIG. 5. Variation of $R_{opt}$ and $W_{opt}$ with $F_{cl}$ shown in (a) and (b) respectively; (c) variation of $s/d_g$ with $F_{cl}$.

We have presented the variation of hopping parameters ($R_{opt}$ and $W_{opt}$) with $F_{cl}$ in Fig. 5(a) and (b) respectively, and the variation of $s/d_g$ (obtained from $T^{-1/2}$ dependence) with $F_{cl}$ in Fig. 5(c). The $R_{opt}$ and $W_{opt}$ values for *types-A* and *B* materials show a rising trend with rising $F_{cl}$. This variation of $R_{opt}$ and $W_{opt}$ are well correlated to the DOS variation with film growth, the lowering of DOS resulting in higher $R_{opt}$ and $W_{opt}$ values. As we reach *type-C* material, $R_{opt}$ and $W_{opt}$ start decreasing abruptly at $F_{cl} \approx 40$, reaching a low plateau level in *type-C* at $F_{cl} \approx 50$. For hopping model to be applicable here, $R_{opt}$ should be greater than or comparable to $d_g$, as the hopping distance has been physically correlated with the effective diameter of Si columns in Si nanostructures. When $R_{opt}$ approaches $d_g$, the hopping process corresponds to near neighbor tunneling, which should give rise to activated behavior of carriers. However, the $R_{opt}$ value in *type-C* is calculated to be 7-9nm (Table I), which is less than the $d_g$ ($\approx$90–100nm), indicating that activated behavior should be evident in this temperature range, which contradicts our observations. Secondly, $\alpha^{-1}$ for highly crystalline polycrystalline material has been reported to be $\approx$257Å.[22] Assuming such a value of $\alpha^{-1}$ for *type-C* material, ($\alpha.R_{opt}$) comes out to be less than one, which precludes the basic assumption of $T^{-1/4}$ hopping mechanism.

Similarly, if we explore the possible microstructural basis of the observed $s/d_g$ variation behavior, we can get an insight into the credibility of Šimánek's model in explaining electrical transport behavior of our material. The $s/d_g$ is low in the *type-A* material, which can be explained by the smaller boundaries of the densely packed small grains. The samples of *types-A* and *B* in this study are all deposited under $R = 1/10$, where non-connecting conglomerate columns are formed having inverted pyramidal shapes. As a result, when the film grows in thickness, density deficit rises with significant void fraction between the columns/ grains. Thus, as we approach *type-B* material, these morphological changes occurring during film growth could result in a rise in the resistive boundary regions and the $s/d_g$ may rise. Nevertheless, the morphological studies of the *type-B* material suggest otherwise, and the onset of conglomeration should result in a lower $s/d_g$ than the derived values, because for any increase in $s$, there is also a significant increase in the $d_g$ of the conducting crystallite conglomerates. However, in *type-C* material, all the samples in this study are deposited under $R = 1/1$ and 1/5. Here a least value of void fraction is seen, resulting in a highly dense material with tightly joined crystallites and straight conglomerate columns. The conglomeration is complete, and a sharp decline in $s/d_g$ signifies the low amount of boundaries compared to the larger width of conducting crystallite conglomerates. Our morphological studies have yielded $s \approx 5\text{--}7$nm and $d_g \approx 90\text{--}100$nm, which leads to $s/d_g \approx 0.05\text{--}0.07$, a value closely matching the value calculated from this model.

Another corroborating evidence for $T^{-1/2}$ dependence in *type-C* material can be derived from the observed value of $T_0$ (Eq. 10), which is a material constant. $T_0$ has been reported to be $1\times10^4\text{--}3\times10^4$ K in highly crystallized nano or microcrystalline Si.[14,16] Value of $T_0$ has been calculated by theoretical method for nc-Si system by Rafiq *et al.*, yielding a value of $1.15\times10^4$K.[17] $T_0$ values in our *type-C* material are $\approx 10^4$K, while in *types-A* and *B*, $T_0 \approx 10^5$ K (Table I). Interestingly, $T^{-1/2}$ dependence of $\sigma_d(T)$ behavior reported for SiF$_4$ based plasma deposited $\mu$c-Si:H material is only seen in fully crystallized (100% crystalline volume fraction) material and not in mixed phase $\mu$c-Si:H system.[16] Therefore, $T^{-1/2}$ dependence is more applicable in *type-C* material. Furthermore, a comparison between the two types of fitting ($T^{-1/2}$ and $T^{-1/4}$) using statistical parameters (like $\chi^2$ or standard deviation) suggests $T^{-1/2}$ to be a better fit to the $\sigma_d(T)$ data of highly crystallized thick samples (*type-C*), but the same cannot be said for the thinner *type-A* samples conclusively. However, a statistical validation of any fit cannot be a reason enough to deem it to be the only explanatory model, but it substantiates the evidence provided by the above-discussed physical parameters. It is, however, instructive to recall the basic assumptions of these two models to examine their plausibility in explaining the observed behavior. The model defining the transport in metal-insulator composite system considers thermally activated carriers and tunneling between nearest neighbors only; and optimization is applied to the product of mobility and number density of charge carriers. In VRH mechanism, density of charge carriers is assumed to be independent of temperature and the percolation paths for the charge carriers are determined by optimizing the mobility. Considering the transport mechanism in metal-insulator composite systems as applied to our heterogeneous $\mu$c-Si:H material, it would be specious to compare the individual constituents of these two systems. Instead, we must take a closer look at the basic process of tunneling as applicable to the film microstructure observed in our material.

The two constituents of $\mu$c-Si:H material, crystalline and disordered phases, have different temperature dependence behavior of $\sigma_d$. It is known that with tunneling conduction, percolation threshold energy does not occur but multiple isolated paths may arise. In a case where the resistance along a path varies greatly, the lowest resistance can dominate the path. In general, such multiple paths having finite width will effectively result in quasi 1-d conduction, thus limiting percolation in such paths. In such a material, an increase in temperature can lower the resistance of a number of unused branches in the conduction network without actually altering the character of the chain of conduction paths. On the other hand, a material in which the resistance does not vary much along a path, an increase in temperature results in cross-linking of such similar paths as in *a*-Si:H which sometimes displays a $T^{-1/4}$ relation. In addition, when conduction occurs along multiple quasi 1-d paths of finite length with weak localization, then $T^{-1/2}$ dependence is seen at low temperatures.[44] This is probably the reason why $T^{-1/2}$ dependence of $\sigma_d$ is seen in those $\mu$c-Si:H films whose microstructure consists of tightly packed large columnar grains without any distinguishable disordered phase (like *type-C* $\mu$c-Si:H), whereas the films having small grains with a fair amount of defect densities in the boundary regions (like *type-A* $\mu$c-Si:H) can show a $T^{-1/4}$ relation.

## VI. CONCLUSIONS

The dark electrical conductivity $\sigma_d(T)$ of $\mu$c-Si:H is intricately linked to the film microstructure. Our study demonstrates that at low temperatures (300–15 K), dark conductivity in highly crystallized undoped $\mu$c-Si:H films can be seen to follow different temperature dependences for different microstructures. The fractional composition of large crystallite grains in $\mu$c-Si:H emerges as a dominant parameter in the study of correlation between film microstructure and its transport properties. Although we have studied the applicability of different transport mechanisms, the choice of an appropriate model was determined by the cross-validation with derived parameter values and film microstructure. A $T^{-1/2}$ dependence of low temperature $\sigma_d(T)$ of highly crystallized $\mu$c-Si:H films (*type-C*) having tightly packed large columnar grains and high percentage fraction of large crystallite grains evince a presence of tunneling of carriers between neighboring conducting crystals, similar to the metal-insulator composite systems. The $T^{-1/4}$ dependence of low temperature $\sigma_d(T)$ in *types-A* and *B* $\mu$c-Si:H films having a low percentage volume fraction of large crystallite grains and low degree of conglomeration can be well addressed by Godet's VRH model that proposes hopping in exponential band tail states. Based on the application of these two different transport models, we have deduced physically rational parameters and material properties of $\mu$c-Si:H material, including DOS values with an exponential distribution near Fermi level, which are in good agreement with film microstructure.